\documentclass{aip-cp}

\usepackage[numbers,sort&compress]{natbib}
\usepackage{rotating}
\usepackage{graphicx}

\usepackage{graphicx}
\usepackage[english]{babel}
\usepackage{listings} 
\begin{document}

\title{Dihadron Production at LHC: 
       BFKL Predictions for Cross Sections 
       and Azimuthal Correlations}

\author[aff1,aff2]{Francesco G. Celiberto \corref{cor1}}
\author[aff3,aff4]{Dmitry Yu. Ivanov}
\eaddress{d-ivanov@math.nsc.ru}
\author[aff2]{Beatrice Murdaca}
\eaddress{beatrice.murdaca@cs.infn.it}
\author[aff1,aff2]{Alessandro Papa}
\eaddress{alessandro.papa@fis.unical.it}

\affil[aff1]{Dipartimento di Fisica, Universit{\`a} della Calabria, 
              Arcavacata di Rende, 87036 Cosenza, Italy}
\affil[aff2]{Istituto Nazionale di Fisica Nucleare, 
              Gruppo Collegato di Cosenza, 
              Arcavacata di Rende, 87036 Cosenza, Italy}
\affil[aff3]{Sobolev Institute of Mathematics, 
              630090 Novosibirsk, Russia}
\affil[aff4]{Novosibirsk State University, 
              630090 Novosibirsk, Russia}

\corresp[cor1]{Corresponding author: 
               francescogiovanni.celiberto@fis.unical.it}

\maketitle

\begin{abstract}

A study of the inclusive production of a pair of hadrons 
(a ``dihadron'' system), having high transverse momenta 
and separated by a large interval of rapidity, is presented. 
This process has much in common with the widely discussed 
Mueller--Navelet jet production and can be also used 
to access the BFKL dynamics at proton colliders. 
Large contributions enhanced by logarithms of energy 
can be resummed in perturbation theory within the BFKL formalism 
in the next-to-leading logarithmic accuracy. 
The experimental study of dihadron production 
would provide with an additional clear channel 
to test the BFKL dynamics. The first theoretical predictions 
for cross sections and azimuthal angle correlations 
of the two hadrons produced with LHC kinematics are presented.

\end{abstract}

\section{INTRODUCTION}

\begin{figure}[t]
 \label{fig:di-hadron}
 \centering
 \includegraphics[scale=0.270]{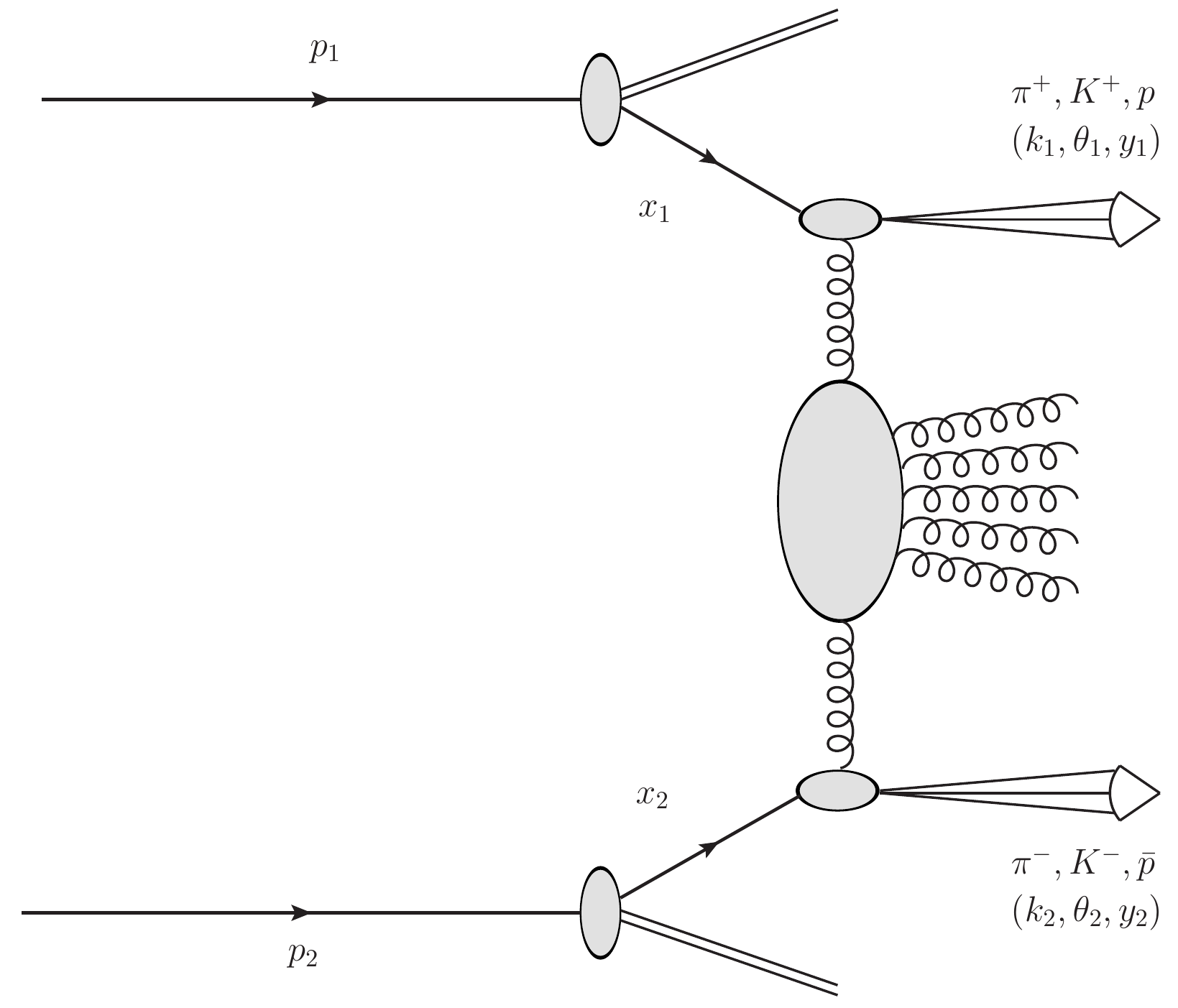}
 \caption{Inclusive dihadron production process in multi-Regge kinematics.}
\end{figure}
The ampleness of data being produced at the Large Hadron Collider 
(LHC) offers us a faultless chance to study the dynamics 
of strong interactions in the high-energy limit.
In this kinematical regime, 
the Balitsky-Fadin-Kuraev-Lipatov 
(BFKL) approach~\cite{Fadin:1975cbKuraev:1976geKuraev:1977fsBalitsky:1978ic} 
represents the most effective tool 
to perform the resummation to all orders of the leading (LLA)
and the next-to-leading terms (NLA) of the QCD perturbative series 
which are heightened by powers of large energy logarithms.
The inclusive production of two jets with high transverse momenta 
and well separated in rapidity, 
known as Mueller--Navelet reaction~\cite{Mueller:1986ey}, 
has been one of the most studied processes in the last years.
For this reaction, the BFKL resummation with NLA accuracy  
relies  on the combination of two ingredients: the NLA Green's function 
of the BFKL equation~\cite{Fadin:1998py,Ciafaloni:1998gs} and the NLA jet 
vertices~\cite{Bartels:2001ge,Bartels:2002yj,Caporale:2011cc,
               Ivanov:2012ms,Colferai:2015zfa}.
In~\cite{Colferai:2010wu,Angioni:2011wj,Caporale:2012ih,
         Ducloue:2013hia,Ducloue:2013bva,Caporale:2013uva,
         Ducloue:2014koa,Caporale:2014gpa,Ducloue:2015jba,
         Celiberto:2015yba,Celiberto:2016ygs,Chachamis:2015crx}  
NLA BFKL predictions of cross sections and azimuthal angle correlations
for the Mueller-Navelet jet process, observables earlier proposed
in~\cite{DelDuca:1993mn,Stirling:1994he,Vera:2006un,Vera:2007kn}, 
were given, showing a very good agreement 
with experimental data at the LHC~\cite{Khachatryan:2016udy}.
In order to further and deeply probe the dynamical mechanisms behind
partonic interactions in the Regge limit, $s\gg |t|$, some other
observables, sensitive to the BFKL dynamics, 
should be considered in the context of the LHC physics program. 
An interesting option, the detection of three and four jets, 
well separated in rapidity from each other, was recently 
suggested in~\cite{Caporale:2015vya,Caporale:2015int} and investigated 
in~\cite{Caporale:2016soq,Caporale:2016zkc,Caporale:2016xku,Celiberto:2016vhn}.
In this paper a novel possibility, {\it i.e.} 
the inclusive production of two charged light hadrons:
$\pi^{\pm}$, $K^{\pm}$, $p$, $\bar p$ 
featuring high transverse momenta and separated by a large rapidity interval,
together with an undetected gluon radiation emission 
is proposed~(see Figure~\ref{fig:di-hadron}). 
For this process, similarly to the Mueller-Navelet reaction, 
the BFKL resummation in the NLA is viable, since NLA expression for 
the vertex describing the production of an identified hadron, 
was obtained in~\cite{hadrons}. 
On one side, hadrons can be detected at the LHC 
at much smaller values of the transverse
momentum than jets, allowing us to explore an additional kinematic range, 
supplementary to the one studied with Mueller--Navelet jets. 
On the other side, this process makes it possible to constrain
not only the parton densities (PDFs) 
for the initial proton, but also 
the parton fragmentation functions (FFs) 
describing the detected hadron in the final state. 
It is known that the inclusion of NLA terms makes a very large effect on the 
theory predictions for the Mueller--Navelet jet cross sections and the jet 
azimuthal angle distributions. Similar features are expected also for our case 
of inclusive dihadron production. This results in a large dependence of 
predictions on the choice of the renormalization scale $\mu_R$ 
and the factorization scale $\mu_F$. Here we will take  
$\mu_R=\mu_F$ and adopt the Brodsky-Lepage-Mackenzie (BLM) 
scheme~\cite{Brodsky:1982gc} for the renormalization scale setting 
as derived in its ``exact'' version in~\cite{Caporale:2015uva}.

\section{DIHADRON PRODUCTION AT THE LHC}

\begin{figure}[h]
\label{fig:hadrons}
\begin{minipage}{.5\textwidth}
\centering
\includegraphics[scale=0.4]{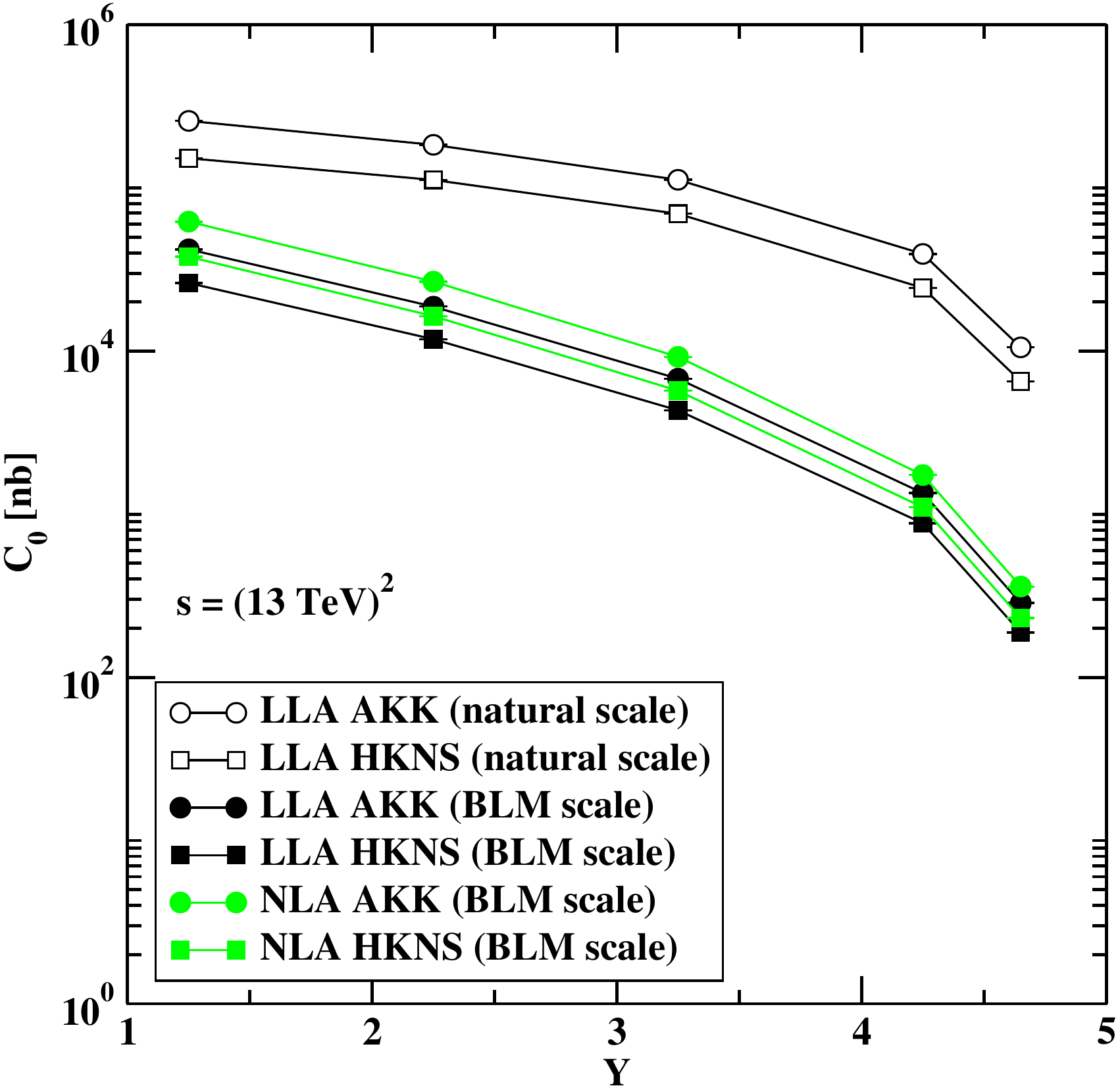}
\includegraphics[scale=0.4]{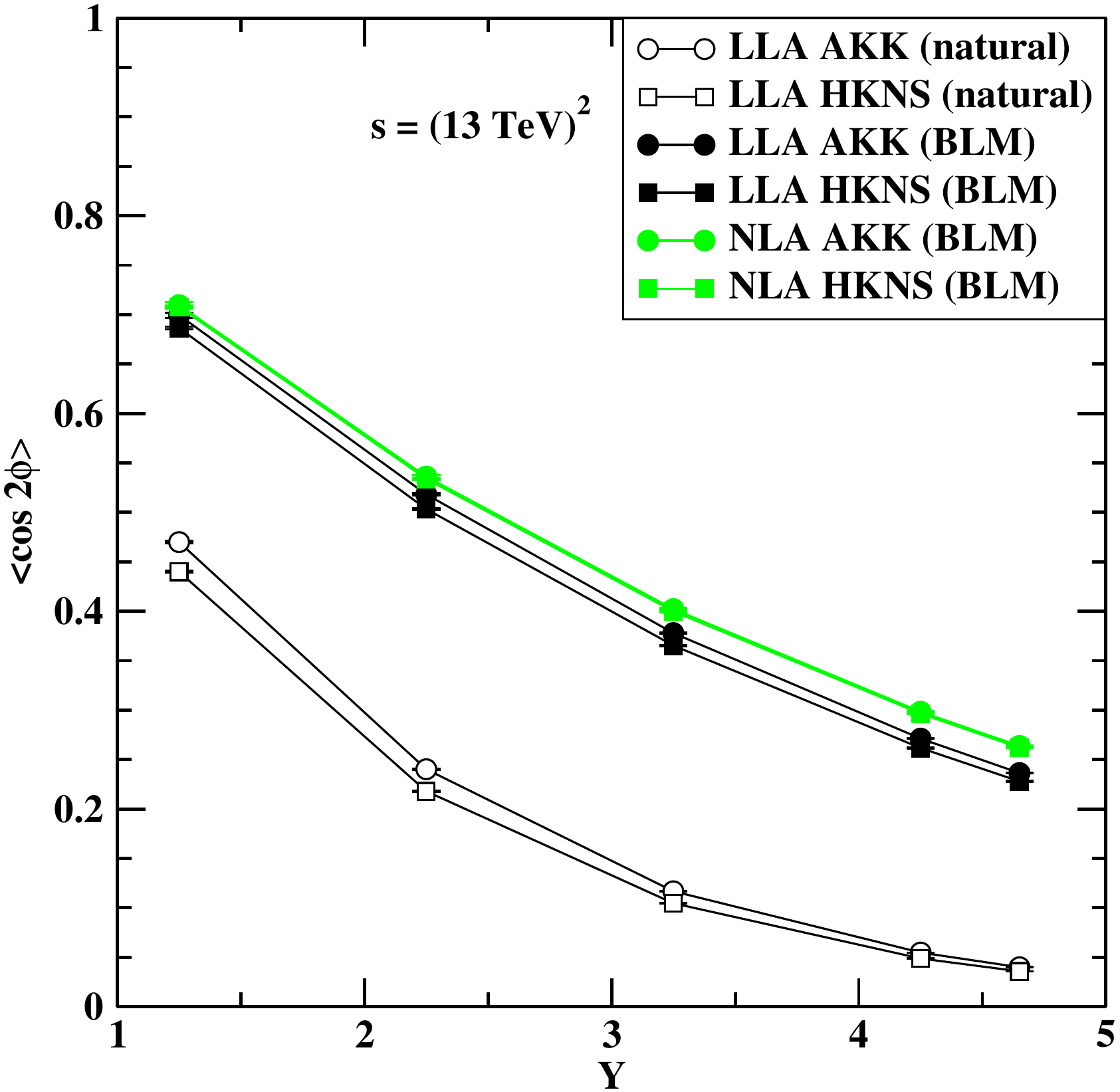}
\end{minipage}
\begin{minipage}{.5\textwidth}
\centering
\includegraphics[scale=0.4]{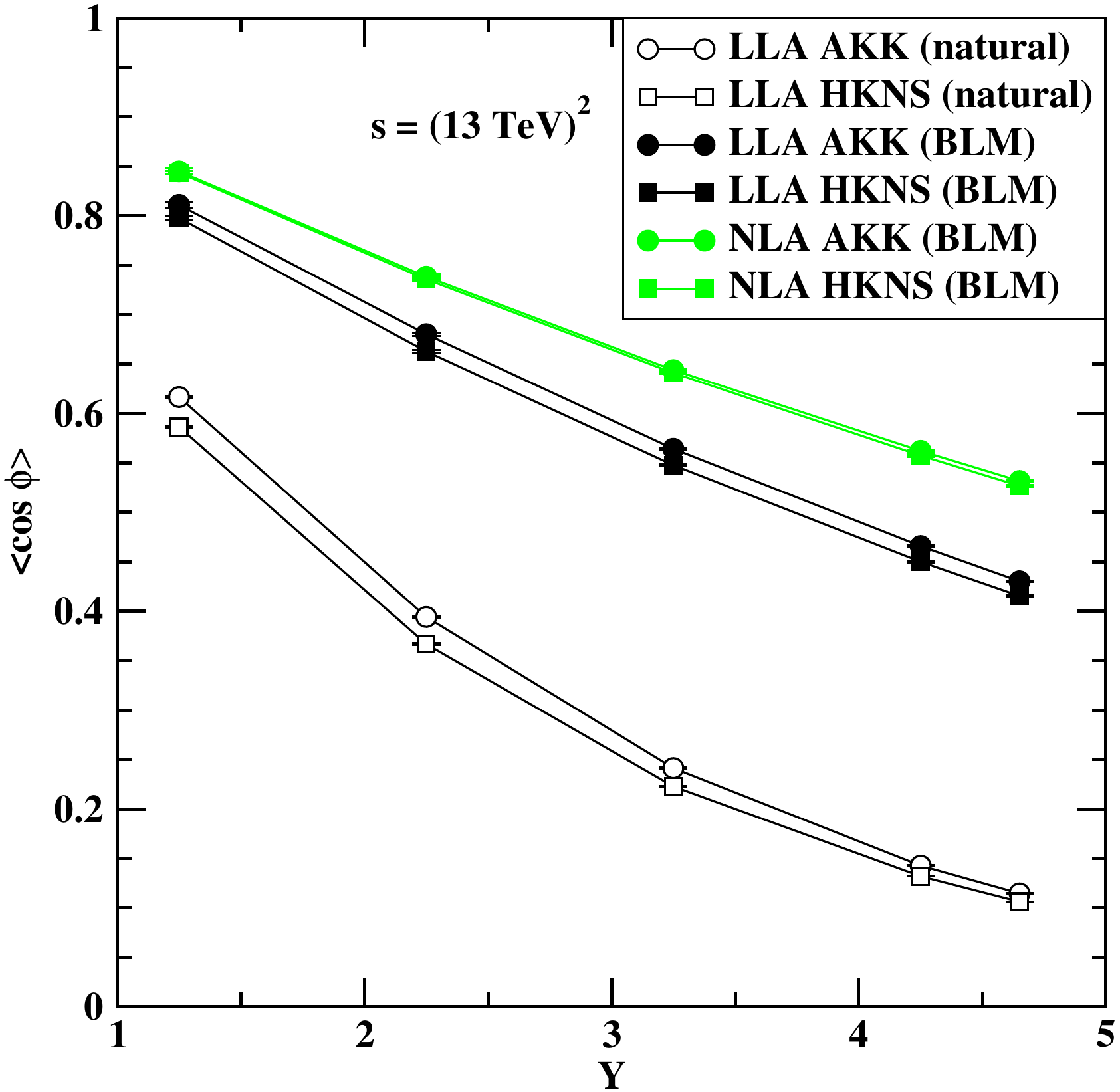}
\includegraphics[scale=0.4]{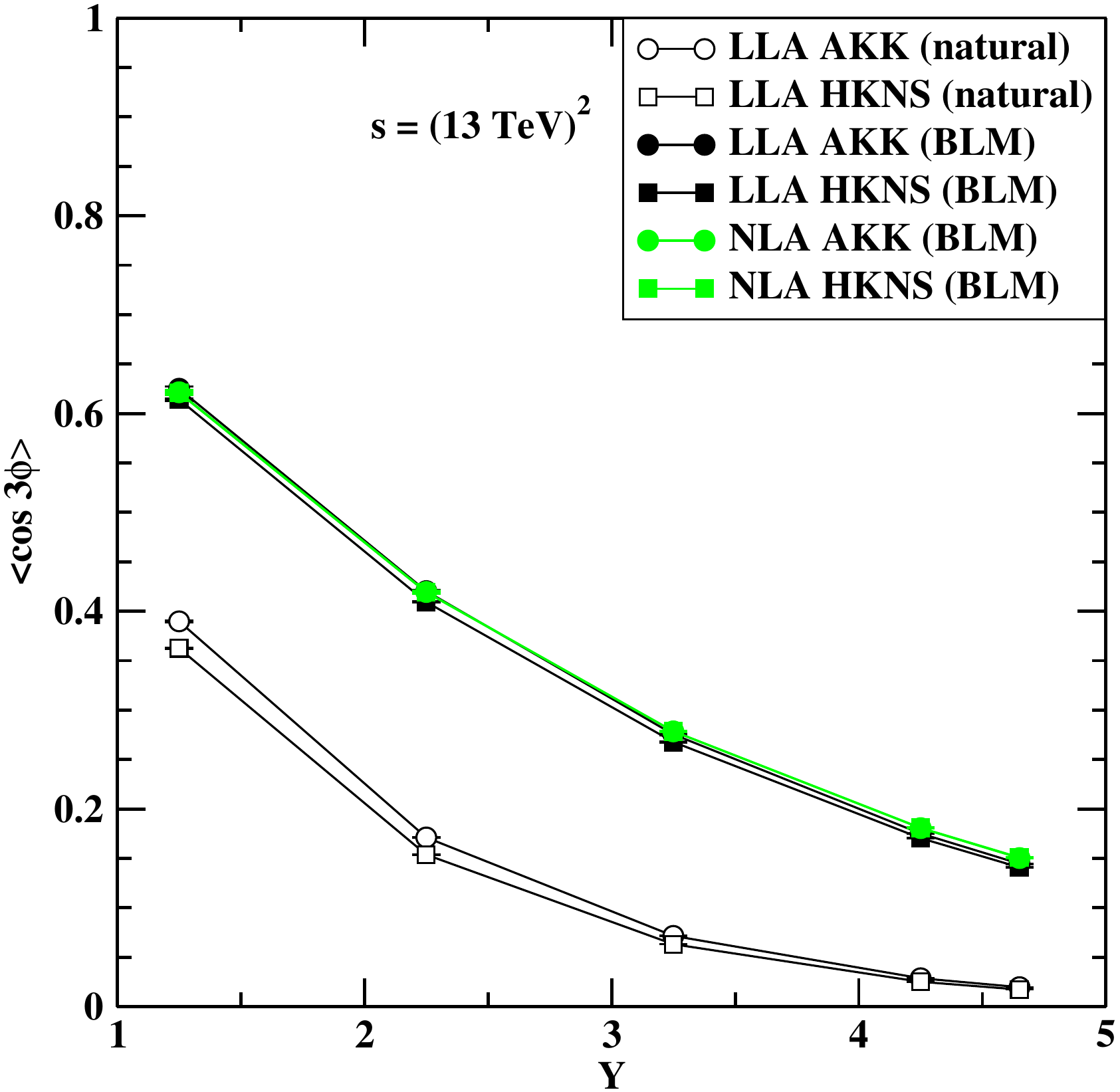}
\end{minipage}
\caption{$Y$ dependence of cross section, 
$R_{10} \equiv \langle \cos \phi \rangle$,
$R_{20} \equiv \langle \cos 2\phi \rangle$, and 
$R_{30} \equiv \langle \cos 3\phi \rangle$
for dihadron production at $\sqrt{s}=13$ TeV. 
See Ref.~\cite{Celiberto:2016hae} 
for predictions of $C_0$ and $R_{10}$ given at larger values of $Y$, 
similar to the ones used in the CMS Mueller--Navelet jets analysis.}
\end{figure}

The process under investigation is the hadroproduction
of a pair of identified hadrons in proton-proton collisions
\begin{eqnarray}
\label{eq:process}
{\rm p}(p_1) + {\rm p}(p_2) \to {\rm h}(k_1, y_1, \phi_1) 
                              + {\rm h}(k_2, y_2, \phi_2) 
                              + {\rm X} \;,
\end{eqnarray}
where the two hadrons are characterized by high transverse momenta, 
$\vec k_1^2\sim \vec k_2^2 \gg \Lambda^2_{\rm QCD}$ 
and large separation in rapidity $Y=y_1-y_2$, 
with $p_1$ and $p_2$ taken as Sudakov vectors.
The differential cross section of the process can be presented as
\begin{equation}
\label{eq:dcs}
\frac{d\sigma}{dy_1dy_2\, d|\vec k_1| \, d|\vec k_2|d\phi_1 d\phi_2}
=
\frac{1}{(2\pi)^2}
\left[
{\cal C}_0+\sum_{n=1}^\infty  2\cos (n\phi ) \, {\cal C}_n \right] \;,
\end{equation}
where $\phi=\phi_1-\phi_2-\pi$, with $\phi_{1,2}$ the two hadrons'
azimuthal angles, while ${\cal C}_0$ gives the total
cross section and the other coefficients ${\cal C}_n$ determine 
the azimuthal angle distribution of the two hadrons. 
In order to match the kinematic cuts 
used by the CMS collaboration, we consider 
the \emph{integrated coefficients}
given by
\begin{equation}
\label{eq:Cm_int}
C_n=
\int_{y_{1,\rm min}}^{y_{1,\rm max}}dy_1
\int_{y_{2,\rm min}}^{y_{2,\rm max}}dy_2
\int_{k_{1,\rm min}}^{\infty}dk_1
\int_{k_{2,\rm min}}^{\infty}dk_2
\, \delta\left(y_1-y_2-Y\right)
\, {\cal C}_n \left(y_1,y_1,k_1,k_2 \right)
\end{equation}
and their ratios $R_{nm}\equiv C_n/C_m$.
For the integrations over rapidities and transverse momenta
we use the limits, 
$y_{1,\rm min}=-y_{2,\rm max}=-2.4$, 
$y_{1,\rm max}=-y_{2,\rm min}=2.4$, 
$k_{1,\rm min}=k_{2,\rm min}=5$ GeV,
which are realistic values 
for the identified hadron detection at LHC.
We use the PDF set MSTW 2008 NLO~\cite{Martin:2009iq} 
with two different NLO parameterizations for hadron FFs:  
AKK~\cite{Albino:2008fy} and HKNS~\cite{Hirai:2007cx}.
Pursuing the goal to stress the potential relevance of 
the process we are proposing, rather than to give a high-precision 
prediction, 
we present our first results neglecting the NLA parts of hadron 
vertices and planning their inclusion as the mail goal 
of a later, more technical publication.
In Figure~\ref{fig:hadrons} the dependence on the
final state rapidity interval $Y$, of the $\phi$-averaged cross section 
$C_0$ and of the ratios $R_{10}$, $R_{20}$, and $R_{30}$
at the center-of-mass energy $\sqrt{s} = 13$ TeV is shown.
Our predictions with the AKK FFs give bigger cross sections, 
whereas the difference between AKK and HKNS 
in the azimuthal correlation momenta is small, 
since the FFs uncertainties are largely canceled out
when we take the ratios $R_{n0}$.
The general features of our predictions are rather 
similar to the respective ones of the Mueller--Navelet jet process. 
Although the BFKL resummation predicts a growth with energy of the partonic 
cross section, the convolution of the latter with proton PDFs 
leads to a decrease with $Y$ of $C_0$. 
The decreasing behavior with $Y$ of the $R_{n0}$ azimuthal ratios 
is due to the increasing amount of undetected parton radiation $\rm X$ 
(see Equation~(\ref{eq:process})) 
in the final state allowed by the growth of the partonic subprocess energy. 
We present our NLA BLM predictions together with the results we 
obtained in LLA, using both BLM and natural scale setting 
($\mu_R^2=\mu_F^2=|\vec k_{1}||\vec k_{2}|$, 
always smaller than the BLM one). 
Plots of Figure~\ref{fig:hadrons} show that LLA results 
at BLM scales lie closer to the NLA BLM ones than LLA results at 
natural scales.

\section{CONCLUSIONS AND OUTLOOK}

In this paper we investigated the dihadron production
process at the LHC at the center-of-mass energy of $13$~TeV, 
giving the first theoretical predictions for cross sections 
and azimuthal angle correlations in the LLA and partial NLA BFKL approach.
We implemented the exact version of the BLM optimization procedure 
in order to make completely vanish the~$\beta_0$-dependence 
in our observables and minimize the size of the NLA corrections. 
We found that our NLA BLM predictions are close 
to the LLA BLM ones, while the LLA calculations at natural scales 
overestimate the total cross section $C_0$ 
and predict a stronger decorrelation for the azimuthal ratios $R_{n0}$. 
The good agreement between LLA and NLA at BLM scales is a direct consequence 
of the small size of the higher-order corrections, 
representing so a clear signal of the reliability of the BLM method. 
However, more accurate analyses are still needed: 
full NLA calculations including next-to-leading order hadron vertices, 
together with the study of larger rapidity intervals in the final state
and considering the effect of a different choice 
for the factorization scale $\mu_F$ 
with respect to the renormalization scale $\mu_R$,
are underway~\cite{Celiberto:2017ptm}.
In view of all these considerations, we encourage experimental collaborations 
to include the study of the dihadron production 
in the program of future analyses at the LHC, 
making use of a new suitable channel to improve our knowledge 
about the dynamics of strong interactions in the Regge limit.

\section{ACKNOWLEDGMENTS}

We thank G.~Safronov and I.~Khmelevskoi for stimulating 
and helpful discussions.
This work was supported in part by the RFBR-15-02-05868.

\nocite{*}
\bibliographystyle{aipnum-cp}%

\end{document}